\documentclass[aps,prd,twocolumn,superscriptaddress,preprintnumbers,floatfix]{revtex4-1}

\usepackage{graphicx}
\usepackage{amsmath}
\usepackage{amssymb}
\usepackage{tikz}
\usepackage{siunitx}
\usepackage{placeins}
\usepackage{color}
\usepackage[colorlinks=true]{hyperref}
\usepackage{standalone}
\usepackage{microtype}
\usepackage{subfigure}
\usepackage{lineno}

\usetikzlibrary{patterns}

\newcommand{\beq}{\begin{equation}}
\newcommand{\eeq}{\end{equation}}

\newcommand{\fgw}{f_{\rm gw}}
\newcommand{\fdotgw}{\dot{f}_{\rm gw}}
\newcommand{\dcc}{LIGO-P1900047}
\newcommand*{\rrpt}[1]{{#1}}
\newcommand*{\rrptd}[1]{{#1}}

\graphicspath{{./fig/}}

\begin{document}


\preprint{\dcc}

\title{Search strategies for long gravitational-wave transients:\\
	 hidden Markov model tracking and seedless clustering}



\author{Sharan Banagiri}
\email[]{banag002@umn.edu}
\affiliation{School of Physics and Astronomy, University of Minnesota, Minneapolis, Minnesota 55455, USA}

\author{Ling Sun}
\email[]{lssun@caltech.edu}
\affiliation{LIGO Laboratory, California Institute of Technology, Pasadena, California 91125, USA}

\author{Michael W. Coughlin}
\affiliation{LIGO Laboratory, California Institute of Technology, Pasadena, California 91125, USA}

\author{Andrew Melatos}
\affiliation{OzGrav, University of Melbourne, Parkville, Victoria 3010, Australia}


\date{\today}

\begin{abstract}

 A number of detections have been made in the past few years of gravitational waves from compact binary coalescences. While there exist well-understood waveform models for signals from compact binary coalescences, many sources of gravitational waves are not well modeled, including potential long-transient signals from a binary neutron star post-merger remnant. Searching for these sources requires robust detection algorithms that make minimal assumptions about any potential signals. In this paper, we compare two unmodeled search schemes for long-transient gravitational waves, operating on cross-power spectrograms. One is an efficient algorithm first implemented for continuous wave searches, based on a hidden Markov model. The other is a seedless clustering method, which has been used in transient gravitational wave analysis in the past. We quantify the performance of both algorithms, including sensitivity and computational cost, by simulating synthetic signals with a special focus on sources like binary neutron star post-merger remnants. We demonstrate that the hidden Markov model tracking is a good option in model-agnostic searches for low signal-to-noise ratio signals. We also show that it can outperform the seedless method for certain categories of signals while also being computationally more efficient.

\end{abstract}


\maketitle



\section{Introduction}

The discoveries of gravitational waves (GW) by Advanced Laser Interferometer Gravitational Wave Observatory (Advanced LIGO) and Advanced Virgo detectors \cite{LIGO2014,Virgo2014} have opened the new window for gravitational wave astrophysics. \rrpt{As of the end of the second observing run (O2)}, LIGO and Virgo have observed multiple binary black hole (BBH) coalescences \cite{Abbott2018catalog} and one binary neutron star (BNS) merger GW170817 \cite{GW170817}, with the latter marking the advent of multi-messenger astronomy \cite{BNSmma}. 

There is also considerable interest in understanding the fate of the BNS merger remnant. In particular, searches have been carried out using O2 interferometric data from LIGO, Virgo, and GEO600 for GW signals from a possible short, intermediate, or long lived remnant of GW170817 with timescales of order of 1\,s, 100--1000\,s and $\gtrsim 1000$\,s respectively \cite{Abbott2017dke, PropertiesGW170817, LVC:2018pmr, vanPutten:2018abw}. Since the nature of the remnant and the exact form of GW emission is unknown, unmodeled searches have played a large role in the analysis. The Stochastic Transient Analysis Multi-detector Pipeline (STAMP), which searches for excess GW power in spectrograms of cross-correlated data, has been employed in both intermediate and long duration searches. These spectrograms are parsed by pattern recognition algorithms for GW signals. Several such algorithms have been proposed in the past; they can be broadly categorized as seed-based and seedless \cite{Burstegard, khan:2009, Thrane:2013bea, Thrane:2010ri}. Seed-based algorithms identify loud seed pixels in the spectrogram (above some threshold) and attempt to grow contiguous clusters from them by adding neighboring pixels. Seedless algorithms \rrptd{pick out} clusters by drawing tracks from some predefined template bank. Since they do not depend on initial loud pixels, they are generally more sensitive than seeded algorithms, especially toward narrowband waveform models albeit at a higher computational cost \cite{Thrane:2013bea, Thrane:2014bma}. Seedless algorithms have to compromise between computational cost and sensitivity to waveform morphology --- for example the implementations of seedless algorithms in \cite{Abbott2017dke, LVC:2018pmr} adopt quadratic fitting in the time-frequency spectrograms.

Here, we apply a hidden Markov model (HMM) tracking algorithm --- first implemented for continuous gravitational wave searches in Refs.~\cite{Suvorova2016,Suvorova2017,Sun2018} --- to cross-correlated data. HMM-based tracking in frequency domain provides accurate estimates of the signal frequency at low signal-to-noise ratio (SNR) when a large number of observational samples are available \cite{Streit1990,Quinn2001}. In the GW context, it was applied in the first observing run (O1) of Advanced LIGO to search for continuous waves from the brightest low-mass X-ray binary, Scorpius X-1 \cite{ScoX1ViterbiO1}. A revised HMM was used to search for signals from a long-lived post-merger remnant of the binary neutron star merger GW170817 \cite{LVC:2018pmr,Sun:2018hmm}. We apply this algorithm to the cross-power maps produced by STAMP and make a quantitative comparison between the performance of HMM and the seedless algorithm.  We demonstrate that HMM can outperform seedless algorithms for specific waveform models, and in particular for models used in post-merger remnant searches \footnote{We focus on the comparison between HMM tracking and seedless clustering because seedless algorithms generally produce better sensitivity than seeded algorithms. We do not discuss seeded algorithms in this paper.}. 

The rest of the paper is organized as follows. In Sec.~\ref{sec:cross-power}, we briefly describe the cross-power maps, the pixel SNR statistic and the detection statistic. In Sec.~\ref{sec:methods}, we describe the two methods being compared --- HMM tracking and the seedless clustering algorithms. In Sec.~\ref{sec:sensitivity_cost}, we compare the detection efficiency and computational cost of the two algorithms for a variety of waveform models, and demonstrate that the HMM tracking generally outperforms in both aspects. A summary of the paper is given in Sec.~\ref{sec:conclusion}.

\section{Cross-power map}
\label{sec:cross-power}

\begin{figure}[h]
	{
		\label{fig:spectogram_sig}
		\scalebox{0.4}{\includegraphics{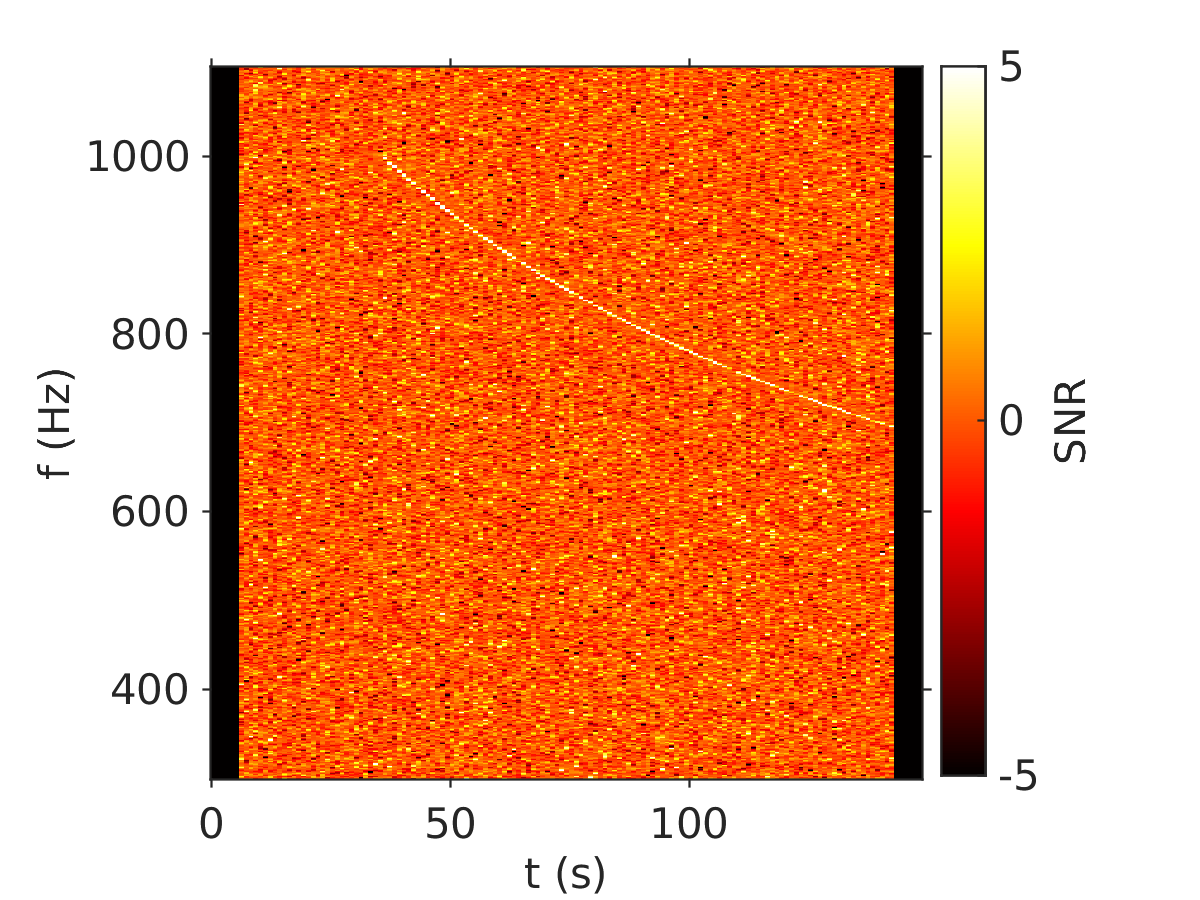}}
	}
	\caption{An example of a cross-power spectrogram in the frequency-time plane, with a loud simulated signal added to it, which is visible as a narrow track. The color of the pixels indicates the SNR.}
	\label{fig:spectrograms}
\end{figure}

Unmodeled transient searches with STAMP are usually done on spectrograms of cross-power. They are constructed by cross-correlating data between two GW detectors in the frequency domain. We follow the definition in Ref.~\cite{Thrane:2010ri} and construct normalized cross-power spectrograms as follows:
\begin{equation}
 \rho (t ;f, \hat{n}) =   \Re  \left [\frac{2 \, \tilde{Q} (t ;f, \hat{n})\, \tilde{s}^*_{I}(t;f) \,  \tilde{s}_{J}(t;f)  }{\left | \tilde{Q} (t ;f, \hat{n}) \right | \sqrt{\frac{1}{2} \, P_I (t ;f) P_J (t;f)}}    \right],	  
\label{pixel snr}
\end{equation}
\rrpt{where $\tilde{s}_{I, J}(t;f)$ is the discrete Fourier transform of data from detector $I, J$ calculated over some segment duration $T$, and $P_{I, J} (t ;f)$ is the \rrptd{noise auto-power \footnote{\rrptd{A common way to estimate the noise auto-power is $P_{I} (t;f) = 2  \overline{|{s}_I (t;f)|^2}  $, where the overline indicates an average over the time segments neighboring $t$ at frequency $f$.}}} in detector $I, J$. Here $\Re$ denotes the real part of a complex number, and $\tilde{Q} (t ;f, \hat{n})$ is a complex filter function which helps ``point" the search in direction of $\hat{n}$ as seen from Earth, given by,} 
\begin{equation}
	\tilde{Q} (t ;f, \hat{n}) = \frac{2 \, \exp{(2 \pi i f \,\hat{n} \cdot \vec{\Delta x_{IJ}}/c )}}{\sum_A F^A_I(t; \hat{n})F^A_J(t; \hat{n})}.
	\label{eq:Qfactor}
\end{equation}
\rrpt{Here $F^A_{I,J}(t; \hat{n})$ is the antenna pattern of detector $I, J$ for polarization \rrptd{$A\in\{+,\times\}$}, $\vec{\Delta x_{IJ}}$ is the distance between the detectors, and $c$ is the speed of light. We point the reader to Ref.~\cite{Thrane:2010ri} for a derivation of Eqns.~(\ref{pixel snr}) and (\ref{eq:Qfactor}).} Since $\rho(t;f, \hat{n})$ has been normalized with the noise PSD, it is called the pixel SNR. Cross-power spectrograms such as in Fig.~\ref{fig:spectrograms} are made by repeating this over many segments of data. 

Pattern recognition or clustering algorithms pick out a cluster of pixels in the spectrogram, representing a possible signal, e.g., one monotonically evolving in frequency. The SNRs of these pixels are then summed up to give the detection statistic. The normalized cluster SNR of the track, $\Gamma_c$ is given by
\begin{equation}
	\label{eqn:gamma}
	\Gamma_c = \frac{1}{\mathcal{N}} \sum_c \rho_i,
\end{equation}
where $i$ indexes over all the pixels in the cluster $c$, and $\mathcal{N}$ is an empirically chosen normalization factor. $\Gamma_c$ plays the role of the detection statistic. The cluster with the largest $\Gamma_c$ will be the trigger for the spectrogram, and the distribution of $\Gamma_c$ is computed for spectrograms containing pure noise to measure the background for the search. A good clustering algorithm finds an optimal cluster which samples as much of a potential signal as possible, without a corresponding increase in the background.

\rrpt{While the existing STAMP analyses have generally used data from two detectors, the formalism could be extended to more detectors. One way would be running the clustering algorithm separately on cross-power spectrograms computed from each pair of detectors, and demanding the triggers be coherent among the pairs. Alternatively we could run the clustering algorithm on the combined spectrogram from all the pairs of detectors ( We refer the readers to section III. E of Ref.~\cite{Thrane:2010ri} for more details).}

\section{Clustering Algorithms}
\label{sec:methods}

In this section, we describe the two clustering algorithms we use here, HMM tracking (Sec.~\ref{sec:hmm}) and seedless clustering (Sec.~\ref{sec:seedless}), using the cross-power spectrograms computed as described in Sec.~\ref{sec:cross-power}.

\subsection{HMM tracking}
\label{sec:hmm}
A HMM is a memoryless, probabilistic state automaton based on a Markov process, composed of the hidden state variable $q(t) \in \{q_1, \cdots, q_{N_Q}\}$ and the measurement variable $o(t)\in \{o_1, \cdots, o_{N_O}\}$ sampled at discrete times $t \in \{t_0, \cdots, t_{N_T}\}$. 
A full description of HMM formulation and the computationally efficient Viterbi algorithm \cite{Viterbi1967} used for solving the HMM can be found in Ref.~\cite{Suvorova2016}. 

We track $q(t)=\fgw(t)$ in a Markov chain, where $\fgw(t)$ is the GW frequency at time $t$. The discrete hidden states $q_i$ are mapped one-to-one to the frequency bins in the cross-power map, with bin size $\Delta f =m/T$, where $m$ is a coarse-graining integer coefficient (see Sec.~\ref{sec:sensitivity}). 
We choose a constant $k$ to satisfy
\begin{equation}
\label{eqn:int_T_drift}
\left|\int_t^{t+T}dt' \fdotgw(t')\right| \leq (k-1) \Delta f,
\end{equation}
for $0\leq t \leq T_{\rm obs}$, where $\fdotgw$ is the first time derivative of the GW signal frequency. 
The HMM emission probability at each discrete time, defined as the likelihood of hidden state $q_i$ being observed in state $o_j$, is given by \cite{Suvorova2016}
\begin{equation}
	L_{o_j q_i} = P [o(t_n)=o_j|q(t_n)=q_i].
\end{equation}
Here we leverage the cross-power pixel SNR in Eq.~\eqref{pixel snr}, and define the emission probability over each time interval $[t,t+T]$ as
\begin{eqnarray}
L_{o(t)q_i} &=& P [o(t)|f_i \leq \fgw(t) \leq f_i+\Delta f]\\
&&\propto \exp[\rho(t;f_i)].
\end{eqnarray}
We also choose $T \leq 100$\,s such that the Earth rotation can be neglected during the interval $[t,t+T]$ in the frequency range of interest. 
The transition probability of $q_i$ from time $t_n$ to $t_{n+1}$ is defined as \cite{Suvorova2016}
\begin{equation}
	\label{eqn:A_matrix}
	A_{q_j q_i} = P [q(t_{n+1})=q_j|q(t_n)=q_i],
\end{equation}
which depends on the signal evolution characteristics. Here we consider a model-agnostic, long-transient signal whose frequency rapidly decreases, e.g., a signal from a binary neutron star post-merger remnant. \rrpt{Assuming that the signal frequency can be approximated by a negatively biased random walk, with frequency change over each segment $T$ uniformly distributed in range $[0, (k-1)m/T]$, i.e., $0\leq \fgw(t_n) -\fgw(t_{n+1}) \leq (k-1)\Delta f$ [see Eqn.~\eqref{eqn:int_T_drift}],} we adopt the transition probabilities 
\begin{equation}
\label{eqn:trans_matrix}
A_{q_{i-j} q_i} = \frac{1}{k},
\end{equation}
with all other entries being zero. In Eq.~(\ref{eqn:trans_matrix}), $j$ takes integer values $0\leq j \leq k-1$. 
We can always adjust Eq.~\eqref{eqn:A_matrix} in searches for other types of signals.
Since we have no independent knowledge of $\fgw$, we choose a uniform prior, viz.
\begin{equation}
\Pi_{q_i} = N_Q^{-1}.
\end{equation}
The probability that a hidden state path $Q=\{q(t_0), \cdots, q(t_{N_T})\}$ gives rise to an observed sequence $O=\{o(t_0), \cdots, o(t_{N_T})\}$ via a Markov chain equals
\begin{equation}
\begin{split}
P(Q|O) = & L_{o(t_{N_T})q(t_{N_T})} A_{q(t_{N_T})q(t_{N_T-1})} \cdots L_{o(t_1)q(t_1)} \\ 
& \times A_{q(t_1)q(t_0)} \Pi_{q(t_0)}.
\end{split}
\end{equation}
The most probable Viterbi path is the maximum a posteriori track, which maximizes $P(Q|O)$. The detection statistic is the cluster SNR as defined by Eq.~\eqref{eqn:gamma} of the optimal Viterbi path. 

\subsection{Seedless clustering}
\label{sec:seedless}
We now briefly describe the seedless clustering algorithm following Refs.~\cite{Thrane:2013bea, Thrane:2014bma}. 
The algorithm attempts to pick out the morphology of a potential signal in the cross-power spectrogram by drawing tracks from a template base. In principle the template could have any possible form. In the case of compact binary coalescence sources for example, one could employ very specific templates drawing upon precise models of GW radiation from them \cite{Coughlin:2014xqa, Coughlin:2014swa}. Yet, in presence of uncertainty about the sources and morphology of astrophysical signals, using quadratic B\'ezier curves is a good tradeoff between sensitivity and computational cost. Quadratic B\'ezier curves provide good sensitivity to many monotonically evolving waveform models, and have been applied in several searches conducted in the past \cite{Abbott2017dke, LVC:2018pmr, Abbott:2017muc}. 

In practice we pick three points (i.e., three pixels) randomly within the spectrogram with the only condition being that the frequency evolution between them be monotonic. The three pixels $N_i = (f_i, t_i)$ are then fit with quadratic curves parametrized by $\xi$:
 \begin{equation}
 	\begin{split}
 		& \begin{bmatrix}
 		f(\xi) \\
 		t(\xi) \\
 	\end{bmatrix}   = (1 - \xi) ^2 N_0 + 2 (1 - \xi) \xi N_1 + \xi^2 N_2.
	 \end{split}
	 \label{eq:bezier}
 \end{equation}
Each B\'ezier template is defined as one choice of $(N_0,N_1,N_2)$, which completely describe a quadratic curve in Eq.~(\ref{eq:bezier}). Usually a total number of templates \mbox{$N_{\rm temp} \sim 10^6$} are used for a single spectrogram \footnote{ \rrpt{ While $N_{temp}$ is tunable parameter, $\mathcal{O}(10^6)$ templates are somewhat of an optimum value. For example using ten times more templates does not result in a substantial increase sensitivity for many waveform models as shown in Ref.~\cite{Thrane:2013bea}} }. The SNR for the cluster obtained from one template, $\Gamma_c$, is defined as the sum of the SNRs of all pixels along the quadratic curve, and is again calculated using Eq.~(\ref{eqn:gamma}). These are the triggers for the search and the loudest trigger for a spectrogram is picked as the prospective signal candidate for further scrutiny. 

\section{Sensitivity and Cost}
\label{sec:sensitivity_cost}

\subsection{Sensitivity analysis}
\label{sec:sensitivity}
In this section, we compare the sensitivities obtained from HMM tracking and seedless clustering. We make comparisons using two different sizes of spectrograms. The ``long duration" spectrograms are 15,000 seconds long, made of short Fourier transforms (SFTs) of 100 seconds of data coarse-grained to 1-Hz frequency resolution \rrpt{(i.e., $m=100$)}. The same configuration of spectrograms has been used in Ref.~\cite{LVC:2018pmr} to search for long-duration post-merger remnant signals. The ``intermediate duration" spectrograms are 500 seconds long with 1\,s SFTs and 1\,Hz bin sizes \rrpt{(i.e., $m=1$)}. Spectrograms of this intermediate size have been used extensively in the past \cite{Abbott:allsky_initialLIGO}, most recently in Refs.~\cite{Abbott:2017muc, Abbott2017dke}. For each configuration, we make a comparison between the two algorithms using (1) Gaussian data recolored to the PSD of the first observing run (O1) of Advanced LIGO, and (2) O2 data from Advanced LIGO Hanford and Livingston detectors, with an unphysical time shift between them \footnote{The time-shift or the time-difference between the data from the detectors is set to be greater than the light travel time and the segment duration. This helps avoid any correlations between true gravitational wave signals in the data.}. The former sets an ideal scenario for comparison, while the latter aims to accurately capture the impact of non-Gaussian, non-stationary artifacts in real interferometer data \footnote{\rrpt{Some existing simulation results using Gaussian data recolored to the O1 PSD have been used in this study to understand
the impact of non-Gaussian, non-stationary noise. Qualitatively, we expect the detection efficiencies obtained from the colored O2 Gaussian data to be comparable to or slightly better than those from the colored O1 Gaussian data. It is difficult to fully quantify the effect of glitches and other non-Gaussian artifacts in real
 data, and the loss of efficiency because of them. It depends on the types of glitches, the waveform model being considered, the efficacy of the glitch rejection algorithms used and the sampling algorithms. We defer any such analysis to a future work.}}.

The long and intermediate duration analyses are described in Secs.~\ref{sec:large-map-sensi} and \ref{sec:small-map-sensi}, respectively, while the waveform models used for simulated signals are described in \ref{sec:sig_models}. For the sake of simplicity and for reducing the computational cost, we fix the sky position of all simulated signals to be the same as GW170817 \cite{BNSmma}, and it is also assumed to be known in the search. All comparisons of sensitivities are made at a false alarm probability (FAP) less than $10^{-4}$. Finally, the search configurations used for HMM tracking and seedless clustering are listed in Table~\ref{tab:search-parameters}, with column 2 and 3 for long and intermediate-duration spectrograms, respectively.

\begin{table}[h]
	\centering
	\setlength{\tabcolsep}{5pt}
	\renewcommand\arraystretch{1.4}
	\begin{tabular}{lll}
		\hline
		\hline
		Parameters & Long duration & Intermediate duration\\
		\hline
		$f$ &30--1800\,Hz &30--1800\,Hz \\
		$T$ & 100\,s & 1\,s \\
		$\Delta f$ & 1\,Hz & 1\,Hz \\
		$T_{\rm obs}$ & 15000\,s & 500\,s\\
		$N_T$ & 150 & 500 \\
		$k$ (HMM) & 10 & 10 \\
		$N_{\rm temp}$ (Seedless)& $10^6$ & $10^6$ \\
		\hline
		\hline
	\end{tabular}
	\caption[parameters]{Search configurations for long and intermediate-duration spectrograms. The last two rows are parameters for HMM or seedless only. From top down, the parameters stand for the frequency band searched, segment duration, frequency resolution, spectrogram duration, HMM configuration constant [see Eq.\eqref{eqn:int_T_drift}], and total number of seedless templates used. }
	\label{tab:search-parameters}
\end{table}

\subsection{Signal models}
\label{sec:sig_models}

\subsubsection{Magnetar model}
For both the long and intermediate-duration spectrograms, we simulate synthetic signals using a neutron star spin-down model, although the search itself is model-agnostic. This model characterizes the gravitational wave radiation from an nonaxisymmetric long-lived post-merger remnant. The remnant might be spinning down due to GW radiation or electromagnetic radiation or some combination thereof. This model has been used in setting limits for post-merger GW emission from GW170817 in both the intermediate-duration and long-duration searches \cite{Abbott2017dke, LVC:2018pmr}. The frequency evolution of the rapidly spinning down signal is given by \cite{magnetar}:
\begin{equation}
\label{eqn:magnetar_freq}
\fgw(t) = \fgw(0)\left(1+\frac{t}{\tau}\right)^{\frac{1}{1-n}},
\end{equation}
where $n$ is the braking index defined via $\fdotgw \propto \fgw^n$, \rrpt{$\tau \propto \fgw^{1-n}/(1-n)$ is the spin-down timescale \cite{LVC:2018pmr}}, and $f_{\rm gw}(0)$ is the starting frequency at reference time $t=0$. The gravitational-wave strain amplitude is given by \cite{magnetar}:
\begin{equation}
\label{eqn:magnetar_waveform}
h_0(t) = \frac{4\pi^2G}{c^4} \frac{I_{zz}\epsilon \fgw^2(0)}{D} \left(1+\frac{t}{\tau}\right)^{\frac{2}{1-n}},
\end{equation}
where $G$ is Newton's gravitational constant, $I_{zz}$ is the principal moment of inertia of the neutron star, $\epsilon$ is its 
equatorial ellipticity, and $D$ is the distance to the source. 

In Table~\ref{tab:inj-para-long-map}, rows 1--2 and 3--4 list the parameters of the synthetic magnetar signals in the long and intermediate-duration analyses, respectively. \rrpt{For all waveform models, detection efficiency at a particular root-sum-squared strain amplitude $h_{\rm rss}$ is defined as the fraction of simulated signals recovered given a FAP of less than $10^{-4}$}. In the frequency domain, $h_{\rm rss}$ is defined as
\begin{equation}
	h_{\rm rss} = \sqrt{2\int_{f_{\rm min}}^{f_{\rm max}} df \left ( |\tilde{h}_+ (f)|^2 + |\tilde{h}_{\times} (f)|^2 \right ) }, 
\end{equation}
where $\tilde{h}_+$ and $\tilde{h}_{\times}$ are strain amplitudes of the waveform in frequency domain for the $+$ and $\times$ polarizations, respectively, and $f_{\rm min}$ and $f_{\rm max}$ are the minimum and maximum frequencies of the waveform in the frequency band being analyzed, respectively.

\subsubsection{Accretion disk instability model}
For intermediate-duration spectrograms, we also test with a different model based on instabilities of accretion disks (ADI) around black holes. These waveforms are parametrized by the mass of the blackhole $M_{BH}$, dimensionless Kerr spin parameter $a^*$, and the fraction of mass forming inhomogeneities in the disk $\eta$. The inhomogeneities are modeled to behave as a binary system and act as a source of gravitational waves. We refer to Refs.~\cite{Abbott:allsky_initialLIGO,VanPuttenADI1,VanPuttenADI2} and the references therein for more details about these waveforms. The last two rows in Tables~\ref{tab:inj-para-long-map} list the parameters of the ADI signals simulated. 

Model spectrograms of the waveforms models used in this paper are shown in the Appendix~\ref{Ap_sec:Spectrograms}

\begin{table}
	\centering
	\setlength{\tabcolsep}{2.5pt}
	\renewcommand\arraystretch{1.4}
	\begin{tabular}{llllll}
		\hline
		\hline
		Model &$\fgw(0)$ (Hz) &$\tau$ (s) &$n$ &Duration (s)& $\cos \iota$\\
		\hline
		magnetar E & 1k & $10^4$& 2.5 & $10^4$ & 1 \\
		magnetar M & 2k & $10^4$ & 2.5 & $10^4$ & 1\\
		magnetar A & 1k & $10^2$& 3 & $10^3$ & 1 \\
		magnetar B & 2k & $10^2$ & 3 & $10^3$ & 1 \\
		\hline
		\hline
		Model &$M_{\rm BH}$ &$a^*$ &$\eta$ &Duration (s)& $f$ (Hz)\\
		\hline
		ADI B &10 $M_{\odot}$ &0.95 &0.2 &9.4 &110--209\\
		ADI C &10 $M_{\odot}$ &0.95 &0.04 &236 &130--251\\
		\hline
		\hline
	\end{tabular}
	\caption{Parameters of the magnetar and ADI models used to generate synthetic signals in Secs.~\ref{sec:large-map-sensi} and \ref{sec:small-map-sensi}. Magnetar models E and M are used for long-duration signal simulations. Magnetar model A and B are used for intermediate-duration signal simulations. The same parameters are used for both Gaussian and time-shifted real interferometer data. The two ADI models are used for intermediate-duration signal simulations, both of which assumes a disk mass of $1.5 M_{\odot}$.}
	\label{tab:inj-para-long-map}
\end{table}

\subsection{Long duration}
\label{sec:large-map-sensi}
 
\begin{figure*}[!ht]
	\centering
	\subfigure[]
	{
		\label{fig:long-magnetarE}
		\scalebox{0.4}{\includegraphics{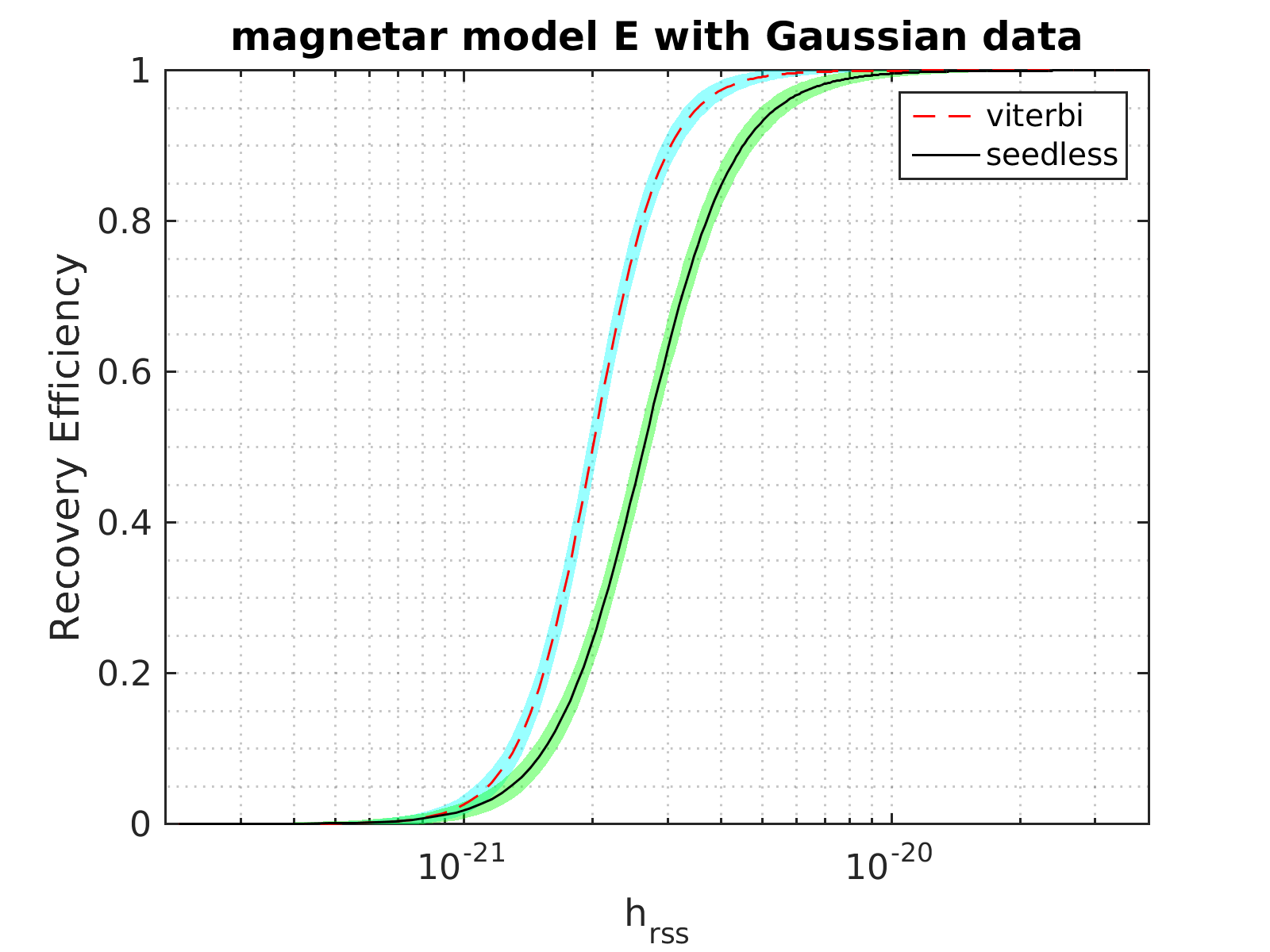}}
	}
	\subfigure[]
	{
		\label{fig:long-magnetarM}
		\scalebox{0.4}{\includegraphics{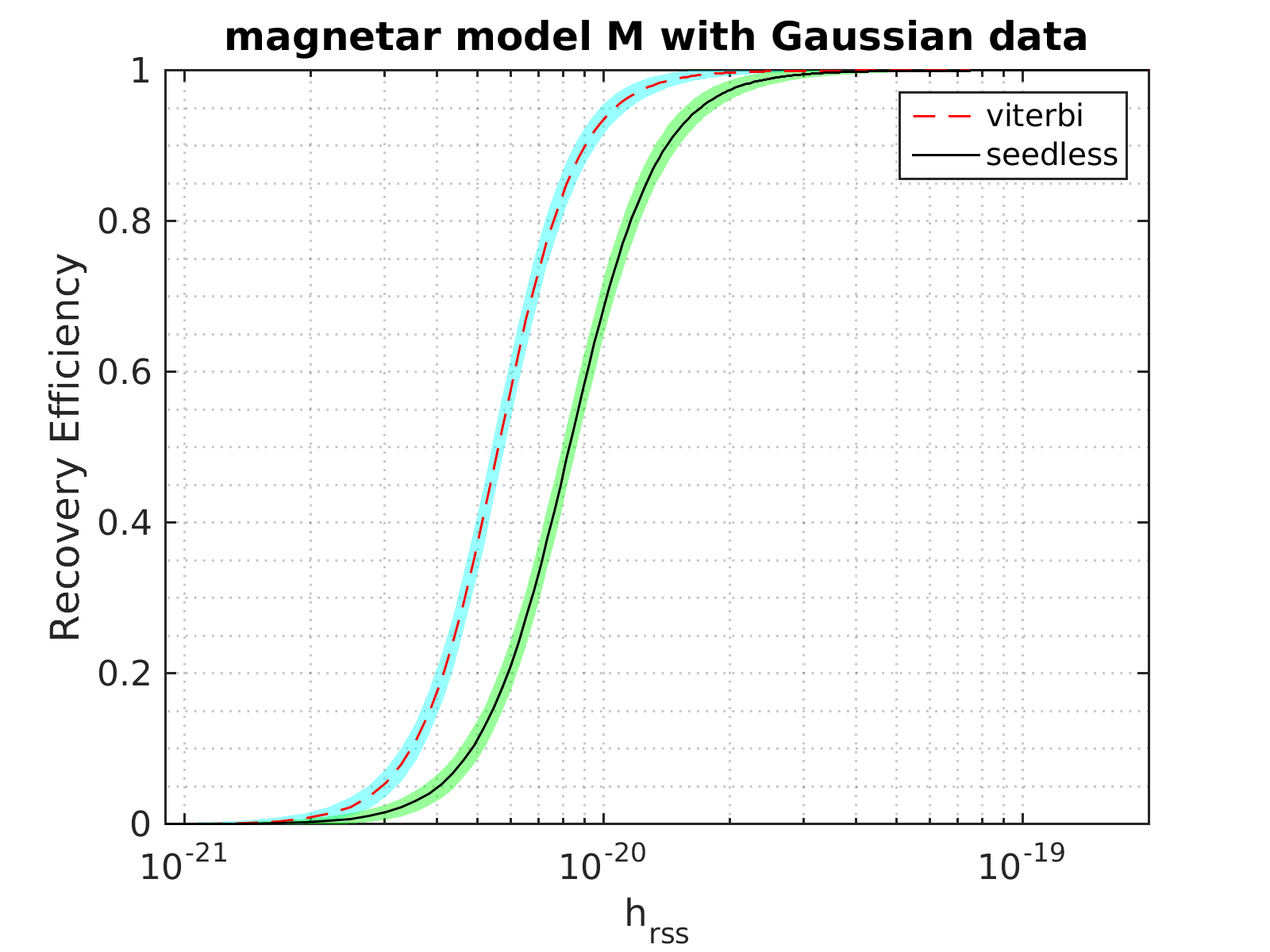}}
	}
	
	{
		\label{fig:long-magnetarE real data}
		\scalebox{0.4}{\includegraphics{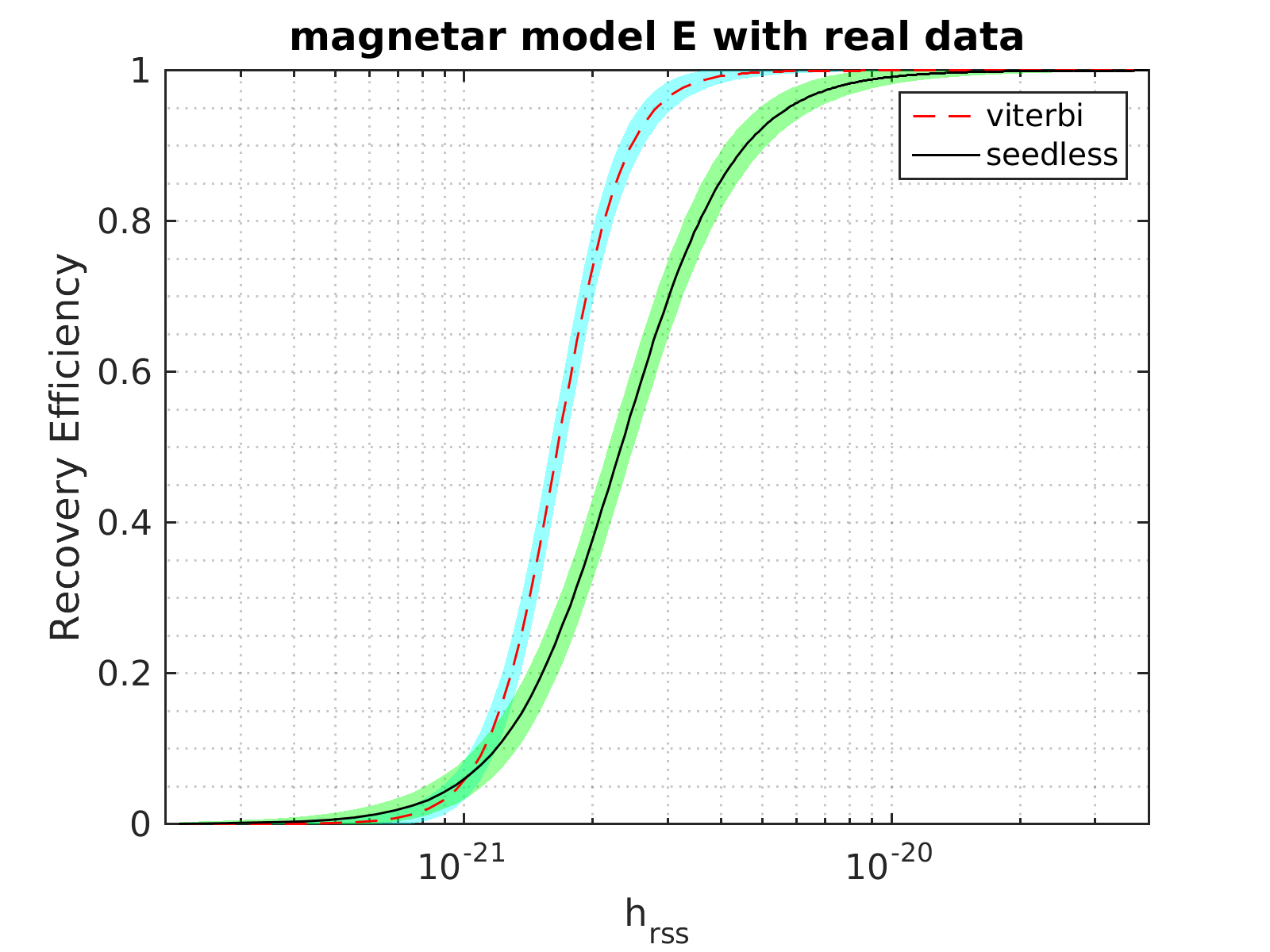}}
	}
	\subfigure[]
	{
		\label{fig:long-magnetarM real data}
		\scalebox{0.4}{\includegraphics{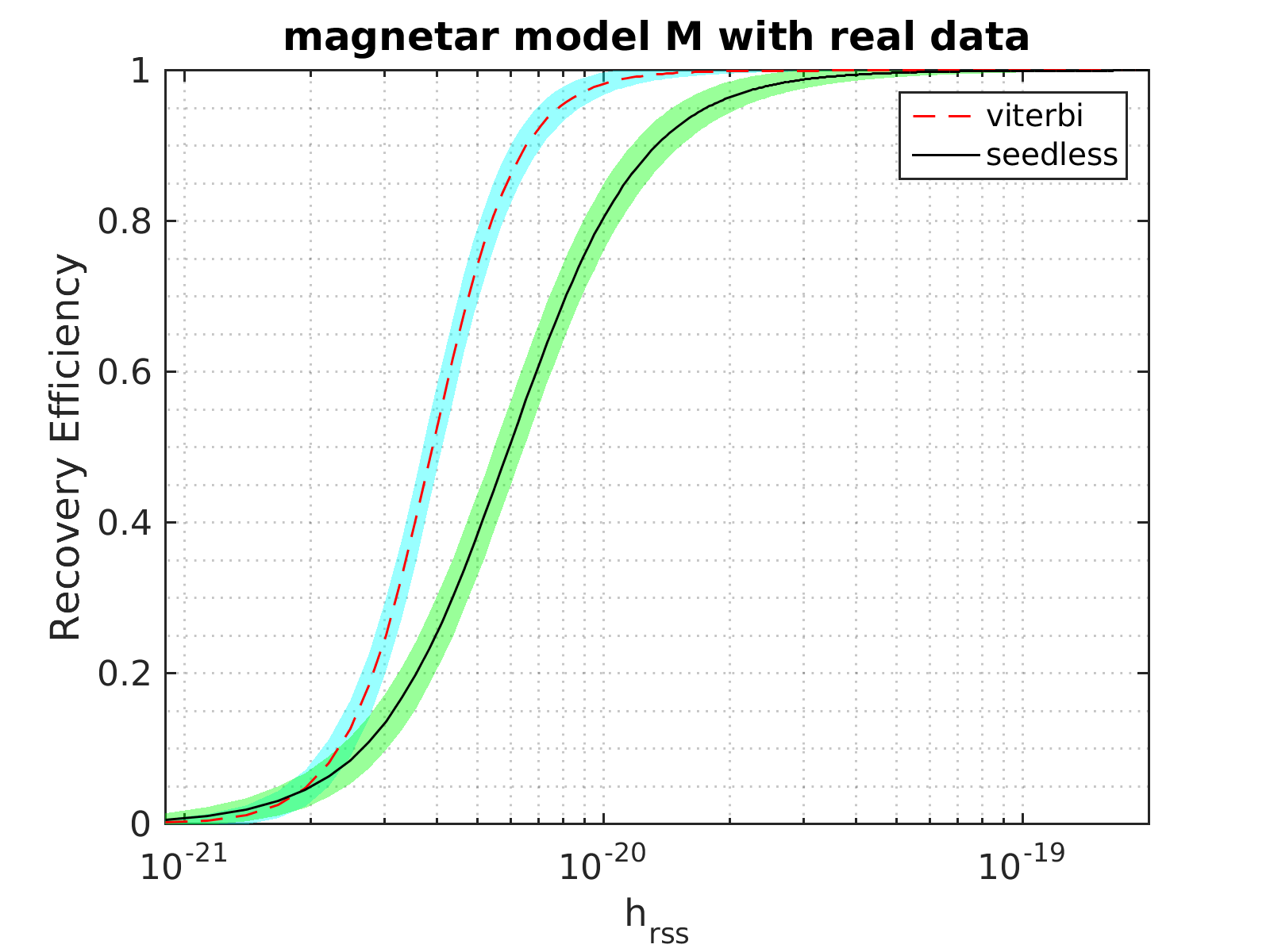}}
	}
	\caption{Detection efficiencies (i.e., the rates of correctly recovering injections) for long-duration simulations based on the magnetar models E and M. \rrpt{The curves shown are generated from sigmoid fits of the discrete injection results.} HMM tracking generally performs better than seedless clustering for these waveforms. The top panels are from simulations in Gaussian noise. The bottom panels are \rrpt{with simulations of the same waveform} injected into time-shifted real O2 data. The results demonstrate that the gain in sensitivity from HMM tracking is not affected by non-Gaussian, non-stationary noise in real interferometric data. We note that the Gaussian data is recolored with O1 noise PSD, and the real data is from the O2 run. The colored regions represent $1 \sigma$ binomial uncertainty in detection efficiency.}
	\label{fig:large-map-msg}
\end{figure*}

\begin{figure*}[ht]
	\centering
	\subfigure[]
	{
		\label{fig:small-magnetar_a}
		\scalebox{0.4}{\includegraphics{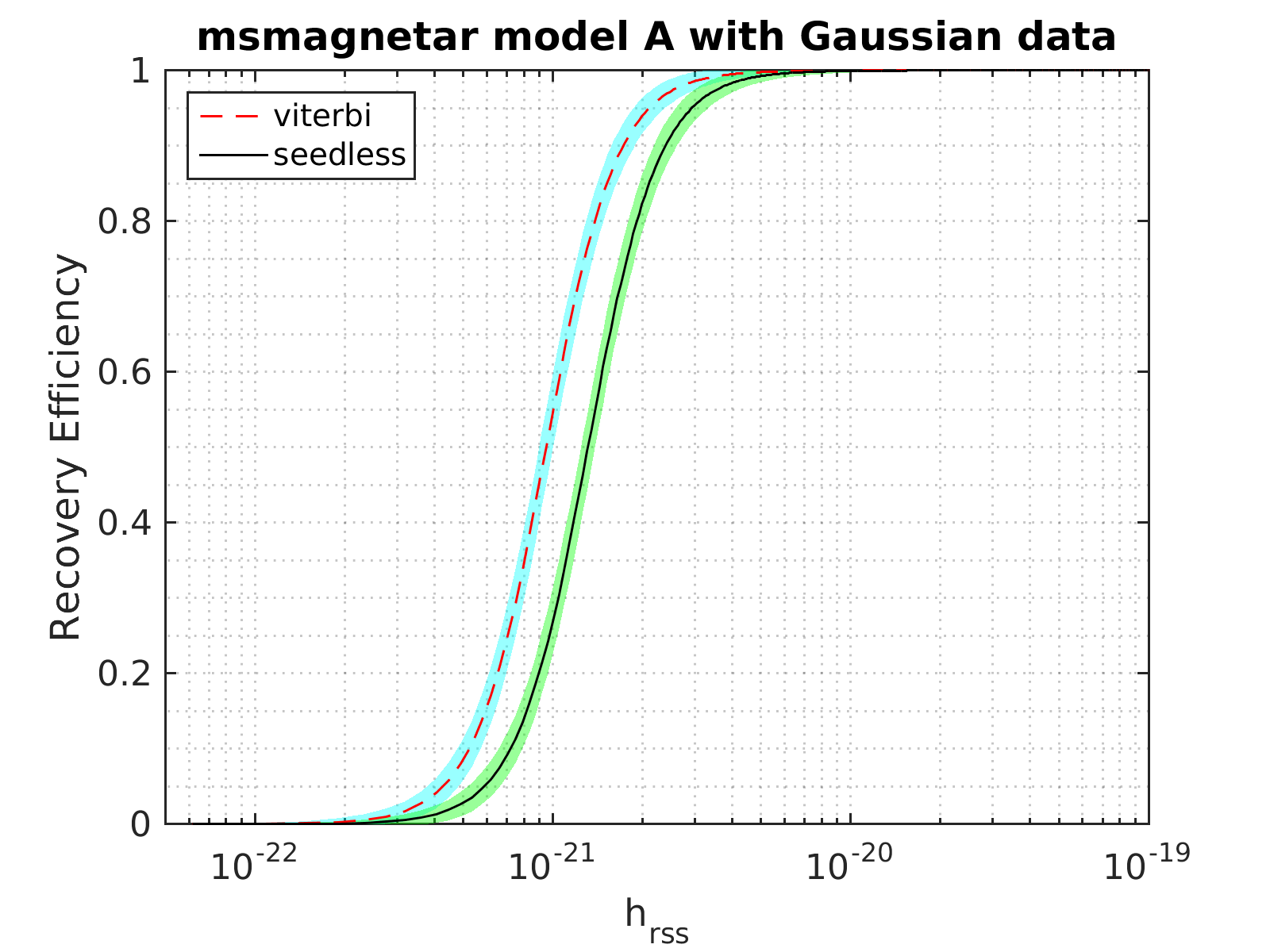}}
	}
	\subfigure[]
	{
		\label{fig:small-magnetar_e}
		\scalebox{0.4}{\includegraphics{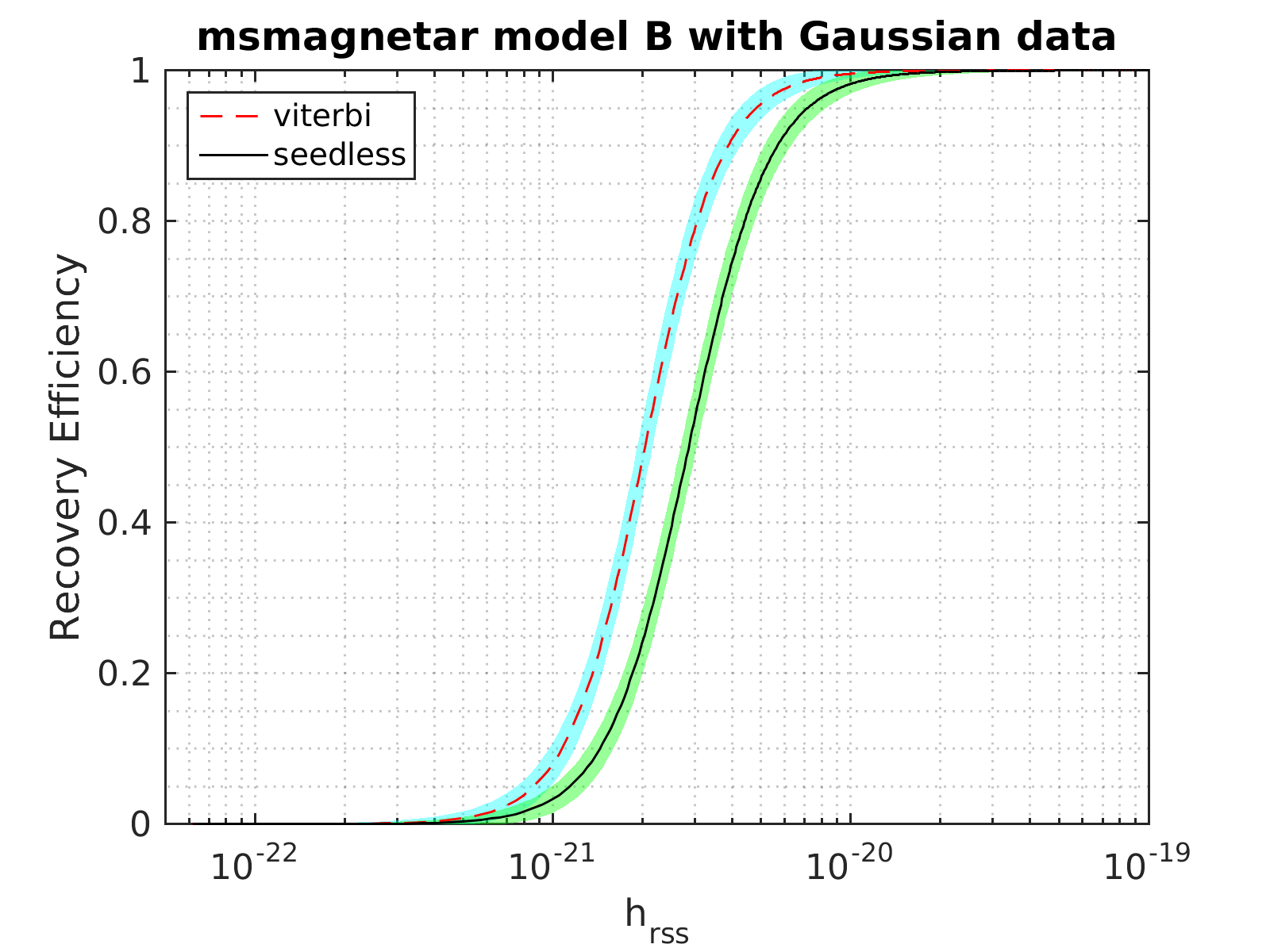}}
	}
	
	\subfigure[]
	{
		\label{fig:small-magnetar_a_real}
		\scalebox{0.4}{\includegraphics{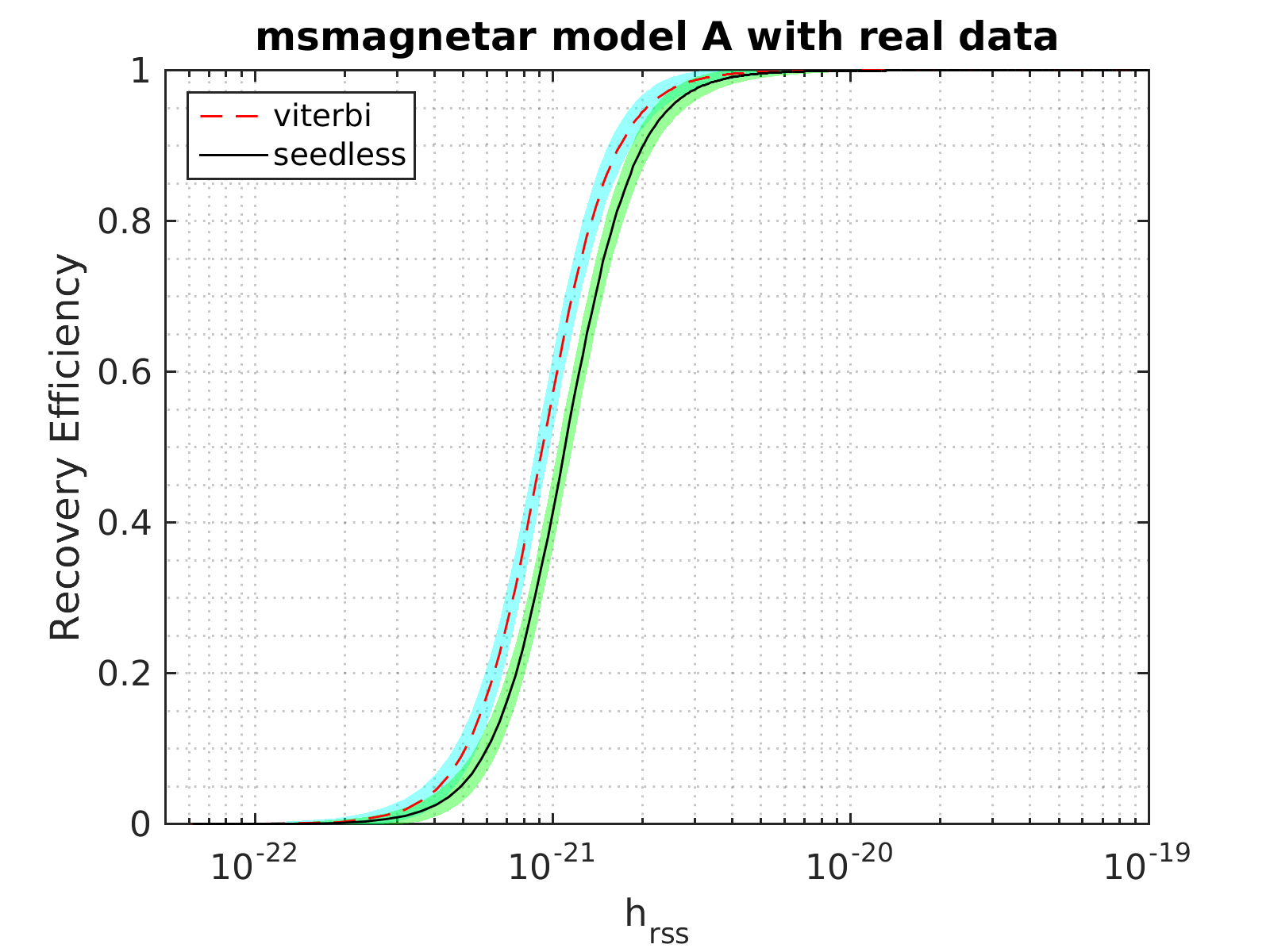}}
	}
	\subfigure[]
	{
		\label{fig:small-magnetar_e_real}
		\scalebox{0.4}{\includegraphics{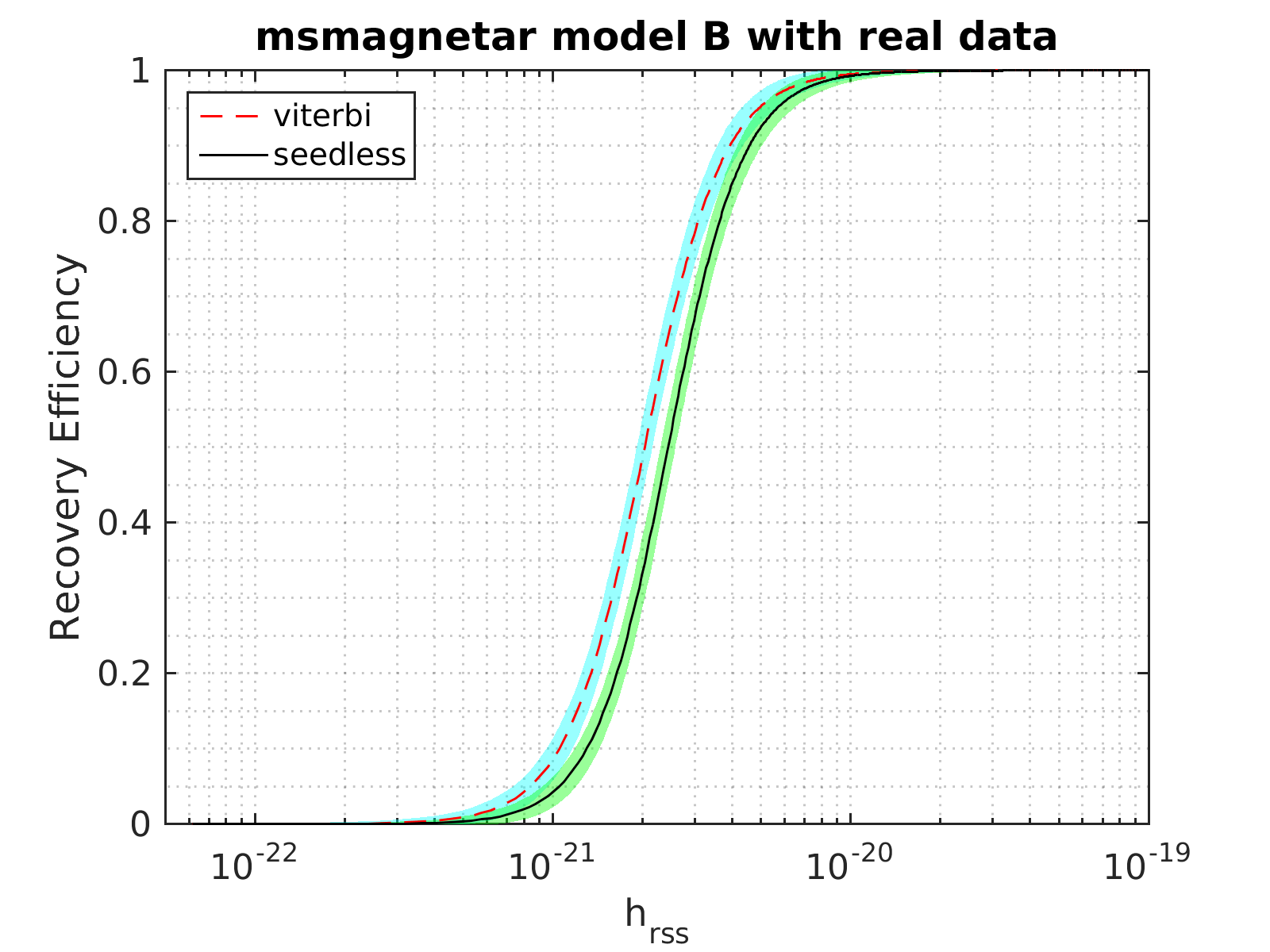}}
	}
	
	\caption{Detection efficiencies for intermediate-duration simulations based on the magnetar models A and B. \rrpt{The curves shown are generated from sigmoid fits of the discrete injection results.} HMM tracking generally performs better than seedless clustering. The top panels are from simulations in Gaussian noise. The bottom panels are \rrpt{with simulations of the same waveform} injected into time-shifted real O2 data. For intermediate-duration signals, the gain in sensitivity from HMM tracking is slightly affected by non-Gaussian, non-stationary noise in real interferometric data. The colored regions represent $1 \sigma$ binomial uncertainty in detection efficiency.}
	\label{fig:short-map-msg}
\end{figure*}

Fig.~\ref{fig:large-map-msg} presents the detection efficiency for HMM tracking and seedless clustering, generated by injecting synthetic signals in Gaussian data recolored with O1 noise PSD (top panels) and real time-shifted O2 data (bottom panels). The dashed and solid curves indicate results from HMM tracking and seedless clustering, respectively. \rrpt{We inject simulated signals at 18 amplitude levels chosen to be uniform in log amplitude. At each amplitude level, 150 simulated signals are used for the Gaussian case and 80 for the real case. These discrete results are then fit to a sigmoid to generate the efficiency curves in Fig.~\ref{fig:large-map-msg}.}

We see that HMM tracking outperforms seedless clustering ($N_{\rm temp} = 10^6$). For example, the strain $h_{\rm rss}$ needed by HMM for an efficiency of 0.9 is lower than seedless by about a factor of two in all cases in  Fig.~\ref{fig:large-map-msg}. A part of the gain comes from the fact that magnetar signal curves are non-quadratic. The quadratic B\'ezier curves used in seedless clustering do not fit the signal well, while the HMM tracking does not assume a particular shape of the signal curve. One can expect that a better template (for instance a template made of cubic B\'ezier curves or from the waveform model itself) will give better sensitivity, albeit at a substantial increase in computational cost or loss in sensitivity to other waveforms.

We also compare the sensitivity obtained here using HMM tracking with cross-power spectrograms to the existing HMM method used in Ref.~\cite{LVC:2018pmr}, which operates on normalized power in SFTs summed over multiple detectors, i.e., $\sum_{X}\tilde{x}^X_i \tilde{x}^{X*}_i $, where $i$ indexes the frequency bins of the normalized SFT $\tilde{x}$, and $X$ indexes the detector \cite{Sun:2018hmm}. With simulations done on time-shifted O2 data, we compute the 90\% sensitive distance, $d^{90\%}$, i.e., the largest distance at which 90\% of the injected signals can be recovered for HMM run on cross-power spectrograms. Using the same moment of inertia of $4.38 \times 10^{38} \, \si{kg.m^2}$ and the maximum possible $\epsilon$ as described in Ref.~\cite{LVC:2018pmr}, we obtain $d^{90\%} \approx 0.5 \, \si{Mpc}$ and $0.2 \, \si{Mpc}$ for magnetar E and M models, respectively. This is done by calculating $h_0$ corresponding to the $h_{\rm rss}$ value required for 90\% efficiency, and using Eqn.~(\ref{eqn:magnetar_waveform}) to convert $h_0$ to the limit on distance. While these are still astrophysically unrealistic distances for a source like GW170817, they are significantly better than the $d^{90\%}$ values ($d^{90\%} \approx 0.064 \, \si{Mpc}$ and $0.035 \, \si{Mpc}$) quoted for similar waveforms in \cite{LVC:2018pmr} using the existing HMM method. 

Note that there are some differences between the simulations in this paper and in Ref.~\cite{LVC:2018pmr}: (1) We use braking index of $n=2.5$ here as opposed to $n=5$ in \cite{LVC:2018pmr}; (2) We use $\cos \iota = 1$ (the inclination of the source) here as opposed to randomized $\cos \iota$ in \cite{LVC:2018pmr}; (3) The FAP in this paper and in \cite{LVC:2018pmr} are $\leq$ $10^{-4}$ and $10^{-2}$, respectively. Although setting $\cos \iota = 1$ can improve $d^{90\%}$ by a factor of 2--3 compared to randomized $\cos \iota$, $d^{90\%}$ for signals with $n=5$ are generally better than $n=2.5$ by a factor of $<2$ \cite{Sun:2018hmm}. Combined with the much more stringent FAP adopted in this paper, these results demonstrate that using HMM operated on cross-correlated spectrograms outperforms its usage on incoherent SFT powers from two detectors. The improvement is not unexpected given that the cross-power SNR statistic demands that the phase difference of the signals between two detectors be consistent with the sky position [see Eqn.~\eqref{pixel snr}], whereas the SFT power spectrograms effectively marginalize over the phase and the sky position \cite{Sun:2018hmm}. A more detailed study of the difference between cross-power and SFT spectrograms is out of the scope of this paper.

\subsection{Intermediate duration}
\label{sec:small-map-sensi}

\begin{figure*}[t]
	\centering
	\subfigure[]
	{
		\label{fig:ADI-B}
		\scalebox{0.4}{\includegraphics{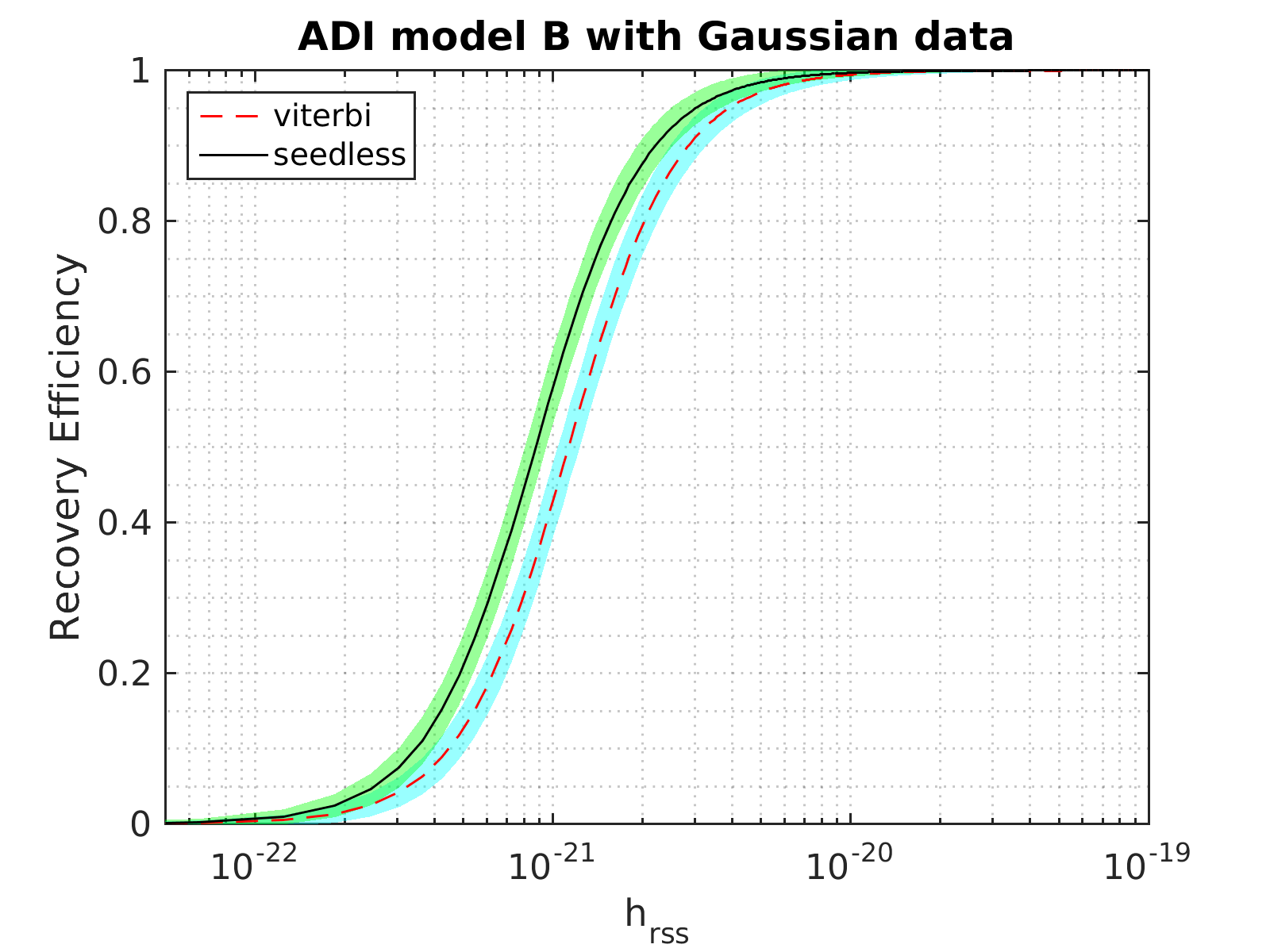}}
	}
        \subfigure[]
	{
		\label{fig:ADI-C}
		\scalebox{0.4}{\includegraphics{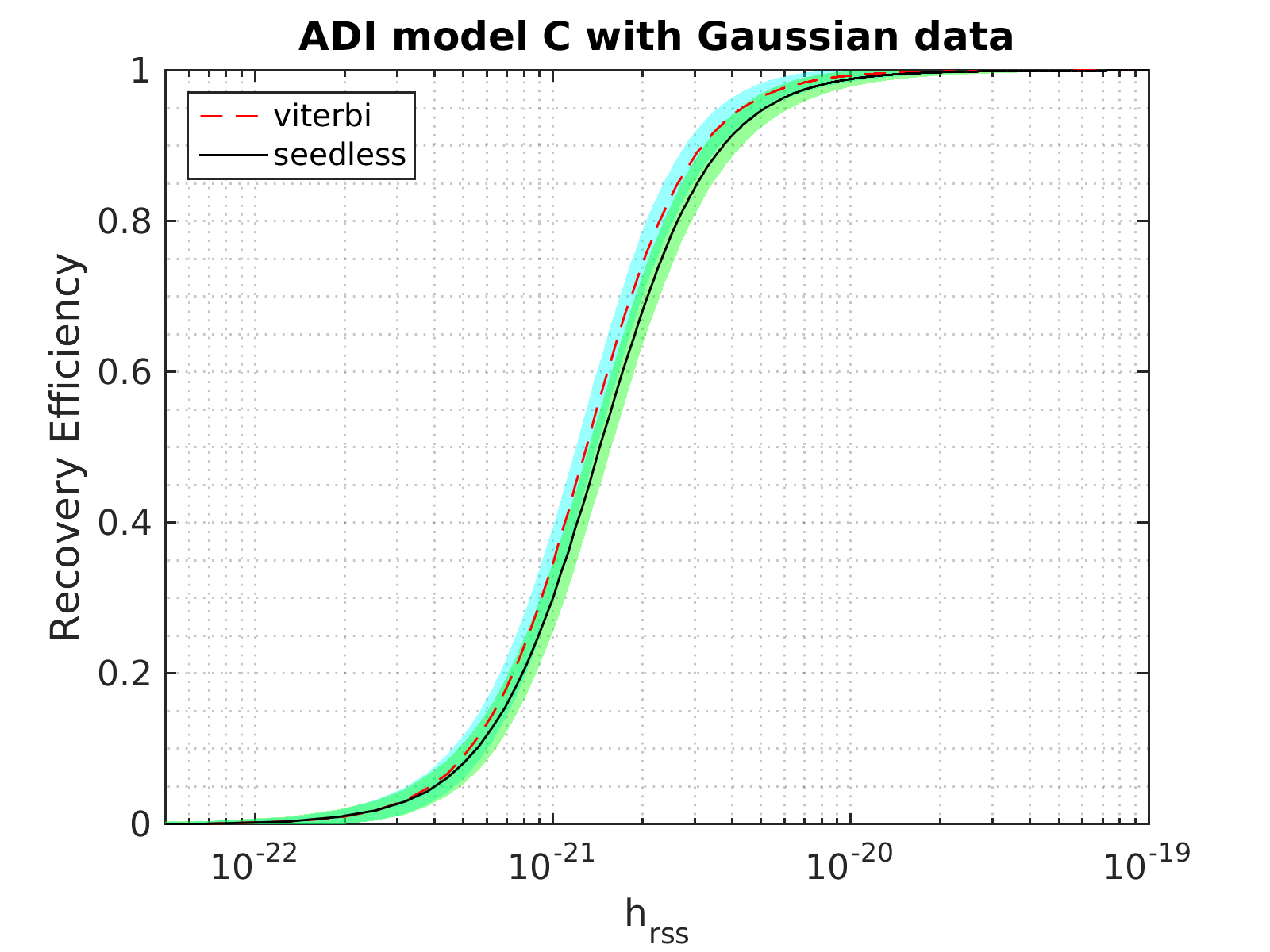}}
	}

	\caption{Detection efficiencies for intermediate-duration simulations based on the ADI models B and C in Gaussian noise. \rrpt{The curves shown are generated from sigmoid fits of the discrete injection results.} The performance of two methods is generally comparable. The colored regions represent $1 \sigma$ binomial uncertainty in detection efficiency.}
	\label{fig:small-map-adi}
\end{figure*}

\rrpt{In the intermediate duration search, we inject simulated signals at 18 amplitude levels for the magnetar model and at 22 amplitude levels for the ADI model, uniform in log amplitude. At each amplitude level, 100 simulated signals are injected. Discrete results are fit to a sigmoid to generate the efficiency curves in Fig.~\ref{fig:short-map-msg} and Fig.~\ref{fig:small-map-adi}.}

Fig.~\ref{fig:short-map-msg} shows the relative performance of HMM and seedless clustering using magnetar models A and B, in both recolored Gaussian noise (O1 PSD) and O2 real interferometric data. HMM tracking still outperforms seedless clustering, although in these intermediate-duration simulations in read data, the sensitivity gain from HMM is not as significant as the long-duration search. 

The same intermediate-duration search configuration (column 3 in Table~\ref{tab:search-parameters}) is used for recovering \rrpt{accretion-disk instability (ADI)} simulations. For illustration purpose, we perform the ADI simulations in Gaussian noise only. The detection efficiencies for these waveforms are shown in Fig.~\ref{fig:small-map-adi}. For the ADI models, the recovery efficiencies from HMM tracking and seedless clustering are generally comparable. Unlike the magnetar models described above, the ADI signal waveforms are better sampled by quadratic B\'ezier curves, and hence seedless is expected to produce similar sensitivity as to HMM.

\subsection{Computing cost}

\begin{table}
	\centering
	\setlength{\tabcolsep}{3pt}
	\renewcommand\arraystretch{1.4}
	\begin{tabular}{lllll}
		\hline
		\hline
		& $N_{\rm pixel}$ & HMM (s) & seedless (s) & ratio\\
		\hline
		Intermediate & $8.9\times 10^5$ &4 & 145 & 36.25\\
		Long & $2.7\times 10^5$ & 1.6 & 37.8 &23.63\\
		\hline
		\hline
	\end{tabular}
	\caption[]{Total number of pixels and median run time of HMM tracking and seedless clustering for intermediate and long duration spectrograms (over 2500 realizations). We note that the tests were run on a computing cluster with machines containing various intel CPU generations, and the run time depends on the CPU architecture. Hence the improvement ratio in the last column is of more interest in the comparison.}
	\label{tab:run-time}
\end{table}

We have demonstrated that HMM tracking provides detection efficiencies better than or at least comparable to seedless clustering in the parameter space considered here. In this section, we show that HMM tracking significantly outperforms seedless clustering with respect to run time and computational cost, and briefly explain the reason. 

We have tested the run time of both HMM and seedless methods (with identical scenarios and configurations) for both the intermediate and long-duration spectrograms. Over 2500 realizations of the intermediate-duration spectrograms, the median run time of HMM tracking was $\sim 35$ times shorter than seedless clustering with $10^6$ templates (see Table.~\ref{tab:run-time}). A similar test for the long-duration spectrograms produces an improvement of a factor of $\sim 23$ for HMM compared to seedless. The gain is less significant in long-duration spectrograms probably because the coarse-grained long-duration spectrograms consist of a smaller number of pixels, $N_{\rm pixel}$, compared to the intermediate-duration spectrogram (see $N_{\rm pixel}$ in Table.~\ref{tab:run-time}). Although the amount of raw data in the 15,000 second long spectrograms is much larger than the intermediate-duration ones, coarse-graining is generally employed \cite{Thrane:2015wla} to reduce the volume of data being analyzed. Coarse-graining is done by averaging both the cross and auto-power in finer frequency bins to give coarser bins. While averaging reduces the computational cost of analyzing long duration spectrograms, it also leads to a loss of sensitivity. The natural Fourier transform frequency resolution of $1/T=10$\,mHz ($T=100$\,s) in the long-duration spectrograms would require a prohibitively large number of seedless templates for analysis. The efficiency of HMM tracking makes it a promising tool to run deeper searches over spectrograms with finer frequency resolution. 

We now briefly discuss the reason why using HMM shows a significant improvement in computational cost. The HMM tracking uses the dynamic programming algorithm, Viterbi, which reduces the total number of comparisons required to find the optimal path from $N_Q^{N_T+1}$ to $(N_T+1)N_Q^2$ in a spectrogram with $N_{\rm pixel} = N_Q N_T$ \cite{Quinn2001,Suvorova2016}. When matrix $A_{q_j q_i}$ only contains ten non-zero terms along the diagonal, the total number of comparisons reduces to $10 N_{\rm pixel} $. Hence in the intermediate and long duration spectrograms, the total numbers of comparisons are $8.9\times 10^6$ and $2.7\times 10^6$, respectively. As a dynamic process, at each step, the algorithm only records $10N_Q = 1.8\times 10^4$ paths for both configurations, but effectively ensures that the optimal one is kept. This is significantly more efficient than fitting $10^6$ curves and summing up the $N_T$ SNR pixels for each curve as is done in seedless.

\section{Conclusion}
\label{sec:conclusion} 

In this paper, we describe two clustering strategies for long-transient gravitational-wave searches, both operating on pre-calculated, cross-power spectrograms.\rrpt{We conduct a large number of simulations - about $\sim 16000$, and $\sim 21000$ for the long-duration (15000\,s) and intermediate-duration (500\,s) signals respectively -  in both Gaussian noise and real interferometric data}. In the simulations and comparison carried out in this paper, we mainly focus on the magnetar model that is adopted in BNS post-merger remnant and other long-transient searches. We have demonstrated that HMM tracking can produce detection efficiency better than or at least similar to seedless clustering, and reduce the computing cost significantly, based on the same sets of spectrograms. HMM tracking can also be applied to track a variety of signal models in addition to the ones tested above, e.g., non-monotonic signals, by adjusting transition probabilities (see Ref.~\cite{Sun2018}). In addition, a small improvement in the sensitive distance will give a relatively large improvement in sensitive volume. With much lower computational requirements, HMM tracking method can be a good option in unmodeled all-sky searches for long-transient signals. In model-agnostic, computationally challenging searches, HMM tracking can prove to be a superior strategy to parse spectrograms.

The HMM tracking algorithm operated on cross-power spectrograms also outperforms the existing HMM tracking operated on SFT power spectrograms used in previous searches for long-duration BNS post-merger signals \cite{LVC:2018pmr}. This new implementation can serve as a more sensitive and efficient alternative in future analyses of the same kind. Finally, this work might also open the window to otherwise prohibitively expensive all-sky long-duration searches (with $\sim 10^4$-s or longer spectrograms), given the significant reduction in computational cost by using HMM tracking.

\section{Acknowledgments}
We are grateful to Maxime Fays, Rich Ormiston, Stuart Anderson and Vuk Mandic for comments and informative discussions. 
S. B acknowledges support in part by the Hoff Lu Fellowship at the university of Minnesota,  and by NSF grant PHY-1806630. 
L. S is a member of the LIGO Laboratory. 
M. W. C is supported by the David and Ellen Lee Postdoctoral Fellowship at the California Institute of Technology.
The authors are thankful for the computing resources provided by LIGO Laboratory. LIGO was constructed by the California Institute of Technology and Massachusetts Institute of Technology with funding from the National Science Foundation, and operates under cooperative agreement No. PHY--0757058. Advanced LIGO was built under award PHY--0823459. 
The research was also supported by Australian Research Council (ARC) Discovery Project DP170103625 and the ARC Centre of Excellence for Gravitational Wave Discovery CE170100004. 
This paper carries LIGO Document Number \dcc.

\appendix
\section{Spectrograms}
\label{Ap_sec:Spectrograms}

\begin{figure}[!tbh]
 	\includegraphics[scale=0.4]{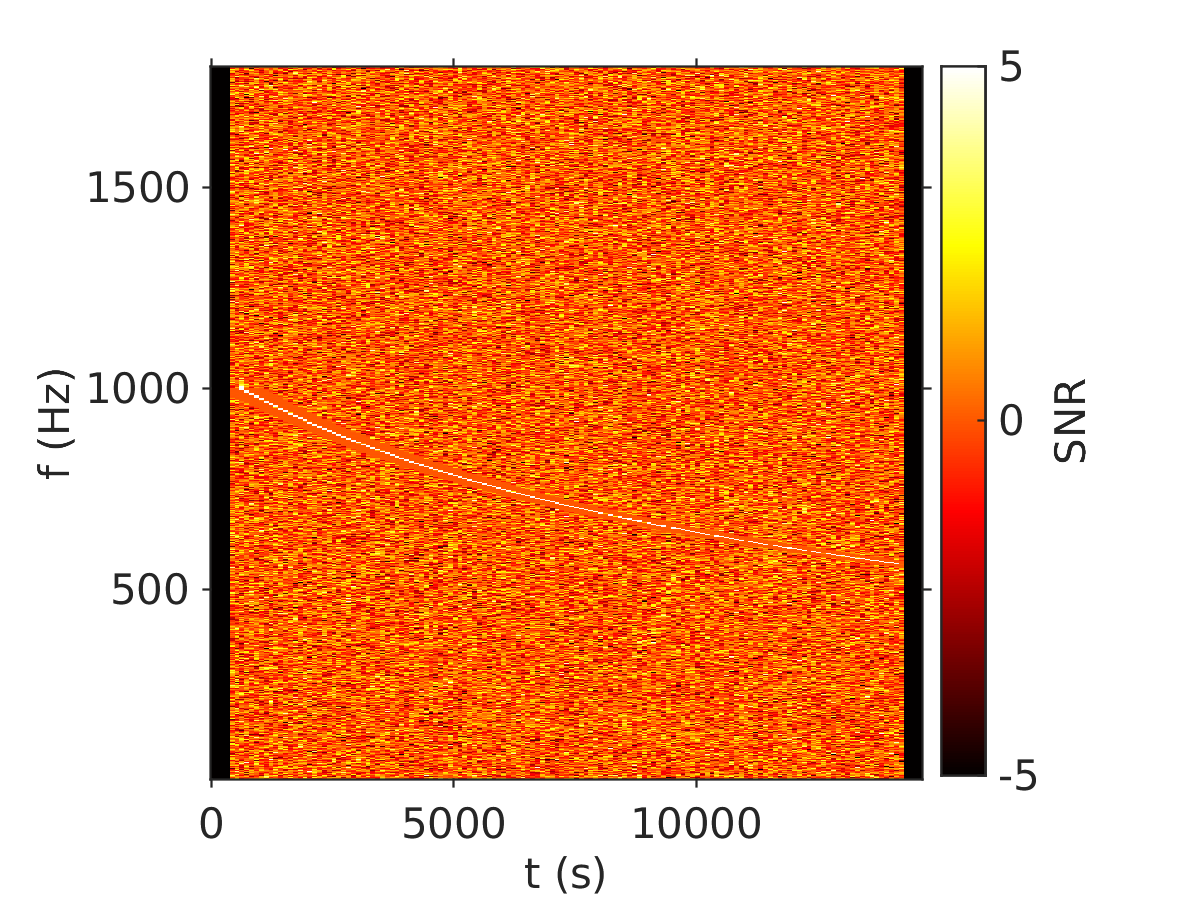}
 	\caption{Sample long-duration spectrogram of the magnetar model E. The spectrogram is for time duration 15000\,s and frequency band 30--1800\,Hz. The signal is visible from about 1000 - 500\,Hz}
 	\label{fig:spectrogram_models_long}
 \end{figure}

\begin{figure}
	\includegraphics[scale=0.4]{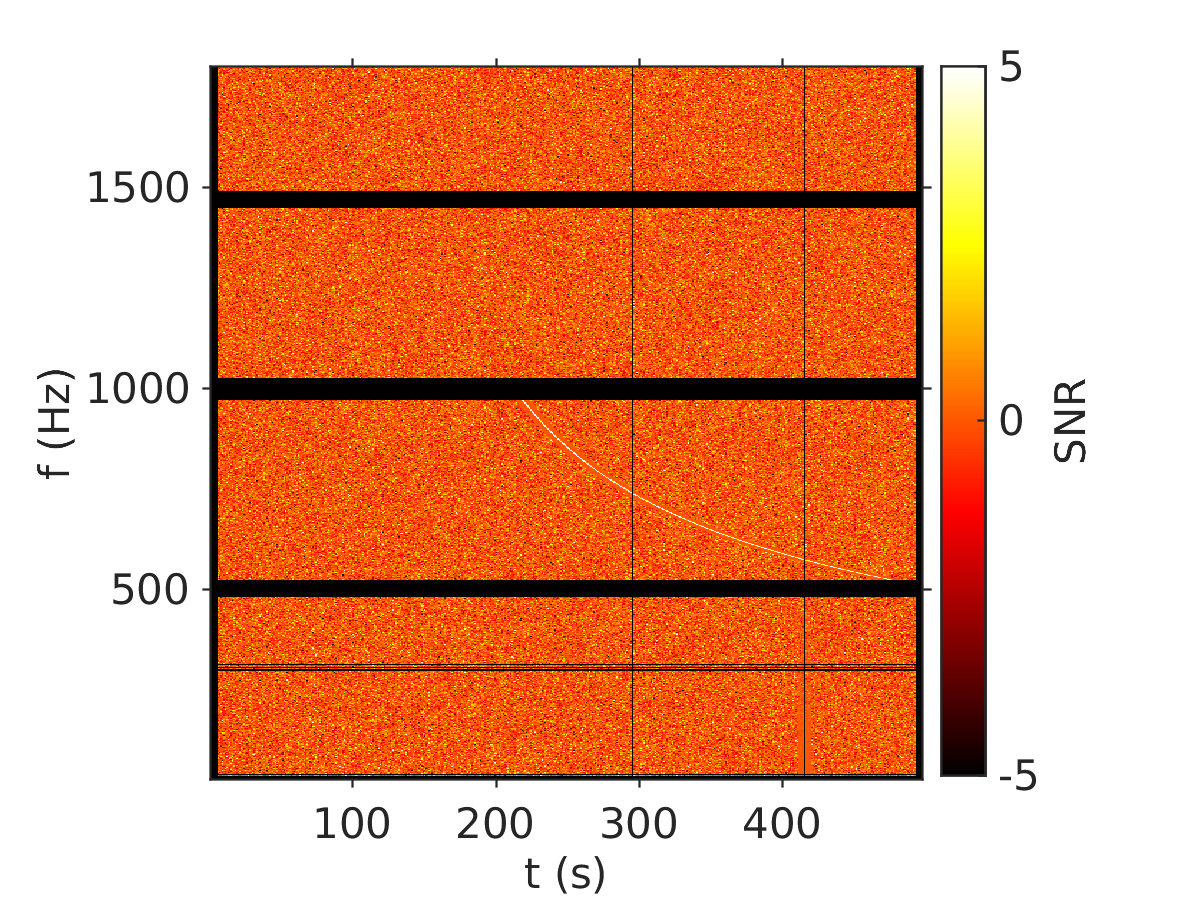}
	\caption{Sample intermediate-duration spectrogram of the magnetar model A. The spectrogram is for time duration 500\,s and frequency band 30--1800\,Hz. The signal is visible from about 1000 - 500\,Hz. The horizontal bars and vertical lines are noisy frequencies which have been notched out or segments which have been vetoed by data-quality cuts. }
	\label{fig:spectrogram_models_mag_a}
\end{figure}

\begin{figure}
	\includegraphics[scale=0.4]{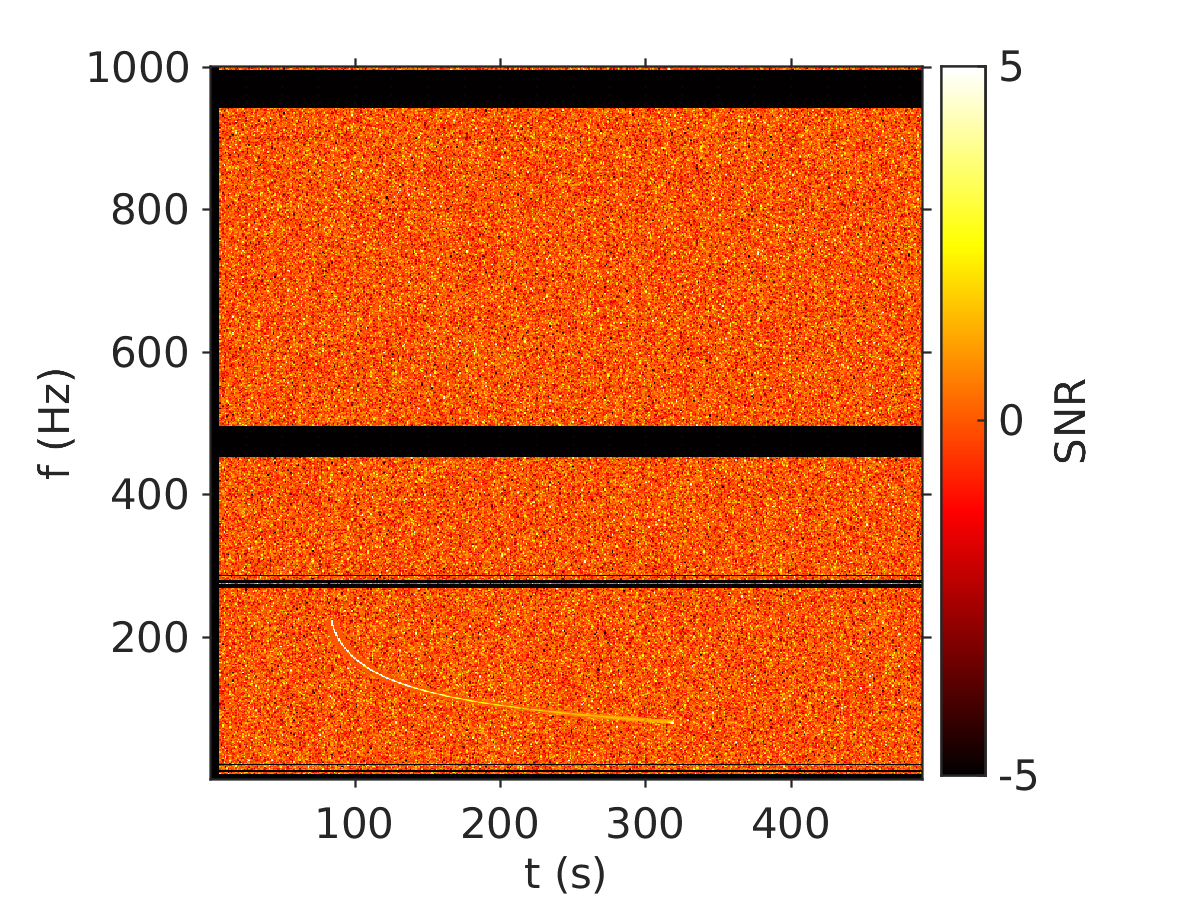}
	\caption{Sample intermediate-duration spectrogram of the ADI model C. The spectrogram is for time duration 500\,s and frequency band 30--1800\,Hz. A sub-band 30--1000\,Hz is displayed here in order to show the injected signal more clearly. The horizontal bars are noisy frequencies which have been notched out. }
	\label{fig:spectrogram_models_adi_C}
\end{figure}

We show sample spectrograms of the waveform models used in this study in Fig.~\ref{fig:spectrogram_models_long} for the long-duration spectrograms and Figs.~\ref{fig:spectrogram_models_mag_a}--\ref{fig:spectrogram_models_adi_C} for the intermediate-duration ones. 
 
\FloatBarrier
\bibliography{stamp-viterbi}

\end{document}